\documentclass[aps,prl,twocolumn,superscriptaddress,longbibliography]{revtex4-2}

\usepackage{amsmath}
\usepackage{amsfonts}
\usepackage{graphicx}
\usepackage{float}
\usepackage{siunitx}
\usepackage{xcolor}
\usepackage{cancel}
\usepackage{mathtools,lipsum, nccmath}

\newcommand{\VEC}[1] {\mathbf{#1}}
\newcommand{\HAT}[1] {\hat{\mathbf{#1}}}
\newcommand{\n}{\VEC{\hat{n}}}

\newcommand{\pdv}[2]{\frac{\partial #1}{\partial #2}}

\begin{document}

\title{Bacterial chemotaxis considering memory effects}

\author{Manuel Mayo}
\affiliation{F\'isica Te\'orica, Universidad de Sevilla, Apartado de Correos 1065, E-41080, Sevilla, Spain}
\affiliation{Departamento de F\'isica, Facultad de Ciencias F\'isicas y Matem\'aticas, Universidad de Chile, Avenida Blanco Encalada 2008, Santiago, Chile}

\author{Rodrigo Soto}
\affiliation{Departamento de F\'isica, Facultad de Ciencias F\'isicas y Matem\'aticas, Universidad de Chile, Avenida Blanco Encalada 2008, Santiago, Chile}

\date{\today}

\begin{abstract} 
Chemotaxis in bacteria such as \textit{E.\ coli} is controlled by the slow methylation of chemoreceptors. As a consequence, intrinsic time and length scales of tens of seconds and hundreds of micrometers emerge, making the Keller--Segel equations invalid when the chemical signal changes on these scales, as occurs in several natural environments. Using a kinetic approach, we show that chemotaxis is described using the concentration field of the protein that controls tumbling in addition to bacterial density. The macroscopic equations for these fields are derived, which describe the nonlocal response.
\end{abstract}

\maketitle


\textit{Introduction.}
Bacterial chemotaxis is usually described by the Keller--Segel (KS) equations, which couple the bacterial density $\rho$ to the ligand concentration $l$ (food, chemoattractant, or chemorepellent)~\cite{keller1970,keller1971model}. In presence of a ligand gradient, bacteria display a chemotactic current that adds to the diffusive one, resulting in the equations
\begin{align} \label{eq.ks}
\pdv{\rho}{t} &= -\nabla\cdot \VEC J,&  \text{ with } \VEC J &= -D\nabla \rho + \mu \rho \nabla l,
\end{align}
where $D$ is the diffusion coefficient and $\mu$ the chemotactic mobility in the linear response regime.
Eqs.~\eqref{eq.ks} describe  chemotaxis at the macroscopic scale, also called the hydrodynamic scale, where gradients are small compared to the  scale of the microscopic agents (bacteria in this case).
As we discuss below, among the various biochemical processes involved in chemotaxis of bacteria like \textit{E. coli}, methylation is the slowest, introducing a new temporal scale of tens of seconds \cite{yuan2012adaptation,tu2013quantitative,zhang2018motor,figueroa20203d} and an associated length scale of hundreds of micrometers \cite{figueroa2020coli,villa2023kinetic}.
These scales are comparable to those appearing in the complex natural environments where bacteria live, such as soils, organs, pores, or even in microscale nutrient patches in sea water (see, for example, \cite{sarkar1994transport,mao2003sensitive,ford2007role,licata2016diffusion,bhattacharjee2019bacterial,blackburn1998microscale,stocker2012marine}).
As a consequence, the simple KS equations are not expected to be valid on those scales and new macroscopic equations need to be derived, which is the purpose of this letter.

Bacteria like \textit{E. coli} move in fluids following the so-called run-and-tumble dynamics~\cite{berg1972chemotaxis}. In the run phase, bacteria move with roughly constant speed and direction, process that is interrupted by rapid reorientations called tumbles. The latter are initiated when there is a reversal in the direction of rotation (from counterclockwise, CCW, to clockwise, CW) of one or multiple flagella~\cite{berg2008coli,taktikos2013motility}. In a simple description, this process of motor switch is governed by the concentration $y(t)$ of the phosphorylated CheY protein (CheY-P)  inside the bacterial body.  Based on the biochemistry of the molecular motor, Tu and Grinstein proposed a model to describe the tumbling process as a two-state activated system~\cite{tu2005white}. In this model, the activation free energy barrier $\Delta G$ and, hence, the transition rate $\nu\sim e^{-\Delta G/kT}$ from the CCW to the CW state depends on the instantaneous concentration $y(t)$. Considering a Taylor expansion of $\Delta G$ and defining the normalized concentration deviation, $X(t ) = [y (t ) - \langle y\rangle]/\sigma_y$, it results
$\nu=\nu_0 e^{\lambda X}$,
where $\sigma_y$ is the standard deviation of $y$ and the dimensionless parameter $\lambda$ measures the sensitivity of the tumbling rate to changes in $X$. 
By tracking \textit{E. coli}, the model was validated, allowing to extract the  parameters: $\nu_0=\SI{0.22}{\second^{-1}}$, $\tau=\SI{19}{\second}$, and $\lambda=1.62$~\cite{figueroa20203d}. As advanced, the memory time $\tau$, which is governed by the methylation process of chemoreceptors, is large compared to the mean run time. 
Associated to it, there is a characteristic memory length $L=V\tau=\SI{500}{\micro\meter}$, where $V= \SI{27}{\micro\meter/\second}$ is the bacterial swim speed measured in Ref.~\cite{figueroa20203d}. 
This relatively large correlation length, comparable to pore sizes in natural environments, generate nonlocal responses that we aim to consider in our description.

Bacteria respond to chemotactic signals by modulating their tumbling rate. As these microorganisms are too small to measure gradients along their body, they integrate the ligand signal on their run, with a chemical pathway that can be modeled by  three principal components: methylation level $m(t)$, kinase activity $a(t)$, and the aforementioned $y(t)$. 
The dynamics of this process can be expressed mathematically using coupled Langevin equations for $a(t) $, $y(t)$, and $m(t)$~\cite{tu2008modeling,tostevin2009mutual,lan2012energy,Ito2015model}. In Ref.~\cite{villa2023kinetic}, by adiabatically eliminating the fast modes of these equations, we showed that the chemotactic coupling can be described by a modification of the Tu and Grinstein dynamical equation for $X$, to incorporate the coupling with the ligand as $ \dot{X}=-(X+b \dot{l})/\tau+\sqrt{2/\tau} \xi$, where in this Langevin equation $\xi$ is a white noise of correlation $\langle\xi(t) \xi(t')\rangle = \delta(t-t')$ and $b$ is the coupling constant to the ligand (positive for attractants and negative for repellers). 
The effects of memory on the temporal response have been successfully described with  these equation~\cite{dufour2014limits,Ito2015model,wong2016role}.
This simple model has been improved in several aspects. Motor adaptation implies that the $\nu$ dependence on $X$ is not exponential~\cite{cluzel2000ultrasensitive,bai2010conformational,yuan2012adaptation,zhang2018motor}. Also, the fluctuation of CheY-P are not small, implying that the equation for $X$ needs to include non-linear coupling terms~\cite{colin2017multiple,keegstra2017phenotypic}. We consider here the most general description for the stochastic dynamics that can accommodate the different models
\begin{align}\label{langevin_eq}
    \dot{X}=-\frac{A(X,l)+B(X,l) \dot{l}}{\tau}+\sqrt{\frac{2}{\tau}} \xi.
\end{align}
Here X is closely related to the concentration of CheY-P, but probably with a change of variable to cast the Langevin equation into one linear in the noise (that is, without multiplicative noise). $A$ and $B$ are dimensionless functions of order one, and we keep $\tau$ as the only relevant slow time scale. We assume that in absence of noise, for any value of $l$ and $\dot l$, there is a single stable fixed point. Also, to ensure the existence of a linear response regime, the scalings $A(X)\sim X$ and $B(X)\sim b$ should be imposed for small $X$. For the tumbling rate we take, $\nu=\nu_0 C(X,l)$, where $C$ is a monotonous increasing function of $X$, normalized such that $C(0)=1$. In what follows, we will work with this general model, but for concrete results we will use the linear model, corresponding to $A(X)=X$, $B(X)=b$, and $C(X)=e^{\lambda X}$.
Finally, Eq.~\eqref{langevin_eq} is coupled to the bacterial motion because $\dot l$ is the rate of change of $l$ in the comoving frame of the swimmer. For a bacterium moving with speed $V$ along the director $\hat{\mathbf{n}}$, it is written as the Lagrangian derivative $\Dot{l}=V \hat{\mathbf{n}} \cdot \nabla l+\frac{\partial l}{\partial t}$.

In this representation, chemotaxis is rationalized as follows. For simplicity, we consider the linear model, but the analysis is analogous for the general case. For a swimmer moving parallel to a chemoattractant gradient ($\hat{\mathbf{n}} \cdot \nabla l>0$ and $b>0$),  $X$ becomes negative on average, reducing 
the tumbling rate, and exactly the opposite results for a swimmer moving against the gradient. The result is a biased run-and-tumble random walk, with longer runs in the direction of the gradient. 
For a chemorepellent ($b<0$), the motion is biased against the gradient.
In absence of memory and fluctuations, $X$ adapts instantaneously to the ligand rate of change, $X=-b\dot l$, resulting in the tumbling rate $\nu=\nu_0 e^{-\lambda b \dot l}$. This approximation or similar ones have been used to describe chemotaxis at the kinetic level~\cite{Schnitzer1993,Chen2003,Lushi2012,Kasyap2012,kasyap2014instability,Bearon2000,tindall2008}.

In Ref.~\cite{villa2023kinetic} we proposed a kinetic equation that incorporated all these elements, which we solved to study the stationary chemotactic mobility and the linear response to signals varying in space and time. The obtained response is nonlocal in space and time, as an effect of  memory. 
The method to build the solutions is not easy to adapt to different geometries and configurations.
A simpler approach is to study the dynamics of slowly-varying fields analog to the KS equations for the density field. With that purpose, we will apply the Chapman--Enskog procedure to the kinetic equation of Ref.~\cite{villa2023kinetic},  which is a systematic method to derive the macroscopic equations for the relevant set of slow fields. 
Other methods have been used to derive macroscopic equations for chemotaxis considering memory~\cite{Chalub2006,XueandOthmer2009,Si2012,Si2014,Xue2015}. However, these equations neglect fluctuations of $X$, which is incorrect, at least for the case of \textit{E.\ coli}. Here, we go beyond this approximation, considering the effects of fluctuations and memory.

\textit{Kinetic description of chemotaxis.}
At the kinetic level, a bacterial suspension is described by the distribution function $f(\VEC r, \HAT n,X,t)$, which is normalized such that the bacterial density is
$\rho(\VEC{r},t)=\int d\HAT{n}\, dX f(\VEC r,\HAT{n},X,t)$.
As shown in Ref.~\cite{villa2023kinetic}, the temporal evolution of $f$ is given by the kinetic equation
\begin{multline}\label{kinetic_eq}
\frac{\partial f}{\partial t}+V \hat{\mathbf{n}} \cdot \nabla f
=\frac{1}{\tau}\left[\frac{\partial^2 f}{\partial X^2}+\frac{\partial(A(X) f)}{\partial X}+\Dot{l} \frac{\partial( B(X)  f)}{\partial X}\right] \\
+\nu_0 C(X)\left[ \int d \hat{\mathbf{n}}^{\prime} w(\hat{\mathbf{n}}^{\prime}\cdot\hat{\mathbf{n}}) f\left(\mathbf{r}, \hat{\mathbf{n}}^{\prime}, X, t\right) -f\right],
\end{multline}
which consider all dynamical effects referred before. To simplify notation, we have suppressed the explicit dependence on $l$ of the coupling functions. The second term in the left hand side describes the streaming of bacteria at velocity $V\HAT n$, the first bracket on the right hand side is the Fokker--Planck term associated to the Langevin equation \eqref{langevin_eq}, describing the evolution of $X$ and, finally, the term on the second line accounts for the tumbling process in the form of a Lorentz term. 
Specifically, the integral term gives the gain rate of bacteria with director $\HAT n$ after a swimmer with director $\HAT{n}'$ made a tumble, and the second term is the loss rate due to tumbling of swimmers having a director  $\HAT n$. 
Both the gain and loss terms are proportional to the tumbling rate 
and when a tumble takes place, the new director is chosen with a probability $w$.
The results of this letter do not depend on the full expression of $w$, but only on its first moment
$\alpha_1 =  \int d \hat{\mathbf{n}}^{\prime} w(\hat{\mathbf{n}}^{\prime}\cdot\hat{\mathbf{n}}) \hat{\mathbf{n}}^{\prime}\cdot\hat{\mathbf{n}}$.
For an  isotropic tumbling, $w=1/\Omega_d$, corresponding to taking the new director uniformly distributed in the unit sphere $\Omega_d$ in $d$ dimensions ($\Omega_2=2\pi$ and $\Omega_3=4\pi$), it results $\alpha_1=0$. For  \textit{E.\ coli}, tumbling is not isotropic, with $\alpha_1\approx0.33$ \cite{berg1972chemotaxis}. Kinetic equations similar to Eq.~\eqref{kinetic_eq} have been previously proposed and used to derive the Keller--Segel equations~\cite{celani2010bacterial} or the compute the currents for stationary ligand gradients~\cite{long2017feedback}.

\textit{Macroscopic equations.}
To derive the macroscopic equation we have to identify which are the slow fields in the system. In this case, that has no spontaneously broken symmetry or critical fields, the only strictly slow fields are associated to conservations~\cite{forster2018hydrodynamic,Chaikin_Lubensky_1995}. Specifically,  bacterial number is the only conservation  and its density $\rho$ is therefore the only slow field. Applying the Chapman--Enskog method to the kinetic equation, it is possible to derive the dynamical equation for $\rho$ at the macroscopic time scale in a formal expansion in spatial gradients (see Ref.\ \cite{Mayo-Soto-PRE} for details). The first nontrivial result appears at second order, obtaining the linear KS model [Eq.~\eqref{eq.ks}] with the diffusion coefficient derived in Ref.\ \cite{villa2020run} and the local chemotactic motility derived in Ref.\ \cite{villa2023kinetic}. To obtain a nonlocal chemotactic response, it would be necessary to continue the Chapman--Enskog method to higher orders in the spatial gradients. This is a standard procedure, although cumbersome, that has been applied for example to  gases \cite{chapman1990mathematical} or rapid granular flows \cite{sela1998hydrodynamic,khalil2014hydrodynamic}, resulting in the Burnett and super-Burnett equations. 
These equations normally have ill-defined boundary conditions that limit their practical applications.

A more fruitful approach was introduced in the study of rapid granular flows, where the inelastic collisions between grains make that mass and momentum are conserved, but energy is not~\cite{andreotti2013granular}. Nevertheless, it was shown that it is possible to apply the Chapman--Enskog procedure to those systems considering as relevant fields, on equal foot, the density, velocity, and energy~\cite{brilliantov2004kinetic,garzo2019granular}. The difference with the standard method is that the energy density evolves also in the fast temporal scale. 

For the eigenvalues of the Fokker--Planck sector in Eq.~\eqref{kinetic_eq},
$\frac{\partial^2 U_n}{\partial X^2}+\frac{\partial(A U_n)}{\partial X} =-\gamma_n U_n$,
it is direct to show that $0=\gamma_0<\gamma_1<\gamma_2\dots$ and that $U_n(X)=u_n(X) \phi(X)$, where $\phi(X)=\phi_0 \exp(-\int^X dX' A(X'))/\Omega_d$, with $\phi_0$ the normalization constant. Finally,  $u_0=1$ and $\int d\n \int dX \phi(X) u_n(X) u_p(X)=\delta_{np}$. The ordering of the eigenvalues and the fact that for  \textit{E.\ coli}, the dimensionless memory time is large, $\hat\tau\equiv\nu_0\tau\approx4.2$, imply that the first CheY-P moment
\begin{align}
\rho_X(\VEC{r},t)=\int d\HAT{n}\, \int dX\, u_1(X) f(\VEC r,\HAT{n},X,t)
\end{align}
can also be considered an approximately slow field for the application of the Chapman--Enskog method. 
The dynamical equations for $\rho$ and $\rho_X$ are obtained taking the appropriate moments of the  kinetic equation~\eqref{kinetic_eq}. That is, this equation is multiplied by 1 or $u_1$, respectively, and integrated over $\HAT n$ and $X$. These moment equations are not closed as fluxes and source terms appear, which depend on higher moments of the distribution. The Chapman--Enskog method precisely allows us to close them as a series expansion in spatial gradients~\cite{chapman1990mathematical,soto2016kinetic}.
 Here we provide the principal elements of the derivation and the complete method is presented in detail in Ref.\ \cite{Mayo-Soto-PRE}. We first assume that the kinetic equation \eqref{kinetic_eq} admits normal solutions, meaning that the spatio-temporal dependence of the distribution function can be enslaved to the evolution of the relevant fields. 
To study the dynamics at the slow hydrodynamic scale, we introduce a formal small parameter $\varepsilon$, which is proportional to the spatial gradients, allowing us to expand the distribution function $f = f^{(0)} + \varepsilon f^{(1)} +\varepsilon^2 f^{(2)} + \cdots$. 
Note that  $\varepsilon$ is placed in front both of the convective term $V\HAT{n}\cdot\nabla f$ and the Lagrangian derivative of the ligand in Eq.\ \eqref{kinetic_eq}.  
Finally, we introduce several time scales $t_0=t, \; t_1=\varepsilon t, \; t_2=\varepsilon^2 t, \; \dots$.
With this, $f =  f[ \n , X | \rho(\VEC r,t_0,t_1,\dots), \rho_X(\VEC r,t_0,t_1,\dots)]$, and the application of the chain rule implies that the temporal derivatives generate a  powers series in $\varepsilon$.

The kinetic equation can now be analyzed order by order in $\varepsilon$. At order $\varepsilon^0$, it is found that, as a consequence of particle conservation, the density field is stationary on the $t_0$ time scale, while $\rho_X$ relaxes on a time scale $\tau$, and the distribution function is given by
\begin{equation}\label{3.19}
 f^{(0)}  = \left[ \rho  + \rho_X u_1(X)  \right]\phi(X),
\end{equation}
where we used the normalization condition that the moments of $f^{(0)}$ are $\rho$ and $\rho_X$.

At first order in $\varepsilon$, it is found that $\rho$ does not evolve on the $t_1$ scale, but $\rho_X$ changes with a rate proportional to $\pdv{l}{t}$.  
Using the linearity and isotropy of the resulting equation, the distribution function can be written as
\begin{multline}\label{3.30}
 f^{(1)}\left[\n,X|\rho,\rho_X\right]  =  \frac{  1 }{  \nu_0}\bigg[   \frac{ \rho M(X) +\rho_X N(X) }{\tau}  \pdv{l}{t} \\ 
 +  V\n \cdot\bigg\{O(X)   \nabla{ \rho}  + P(X)    \nabla{ \rho_X} 
 +\frac{1}{\tau}\left[\rho Q(X) + \rho_X R(X)\right]    \nabla{l} \bigg\}\bigg],
\end{multline}
where the unknown functions satisfy linear equations in  terms of the Fokker--Planck and tumbling operators. For the linear case, similarly to  Refs.\ \cite{villa2020run,villa2023kinetic}, explicit expressions can be obtained as expansions in Hermite polynomials. Details can be found in the Supplemental Material and Ref.~\cite{Mayo-Soto-PRE}.

Finally, the analysis at second order in $\varepsilon$, gives that both  fields evolve in the slow $t_2$ time scale, with irreversible fluxes $\VEC{J}$ and $\VEC{J}_X$ and  a source terms $S_1$ and $\VEC S_2$ that result from the spatial gradients in $f^{(1)}$. Collecting all terms, the hydrodynamic equations that extend the KS model, which are the principal contribution of this letter, are
\begin{align}
    \pdv{\rho}{t} + \nabla \cdot\VEC{J} &= 0 ,\label{complet_rho_eq}\\
   \pdv{\rho_X}{t} + \nabla \cdot \VEC{J}_X &= -\frac{1}{\tau} \left[ \gamma_1\rho_X + \pdv{l}{t} S_1 + \nabla l\cdot \VEC S_2\right],   \label{complet_rhox_eq}
\end{align}
with
\begin{align} 
\mathbf{J} &=    - D_{11} \nabla \rho + D_{12} \nabla\rho_X +(\mu_{11} \rho -\mu_{12} \rho_X) \nabla l, \label{J_rho} \\
   \mathbf{J}_X &=   D_{21} \nabla{\rho} - D_{22} \nabla{\rho_X} -(\mu_{21} \rho - \mu_{22} \rho_X ) \nabla{ l} ,\label{J_X}\\
    S_1 &= g_1\rho + g_2\rho_X + (g_3 \rho +g_4\rho_X)\pdv{l}{t},  \\
\VEC{S}_2 &= g_5 \nabla \rho + g_6\nabla\rho_X + (g_7\rho+g_8\rho_X)\nabla l.
\end{align}
The constants $g_i$, and the diffusion and mobility coefficients are expressed as integrals of $M, N, \dots R$, given in the Supplemental Material and Ref.~\cite{Mayo-Soto-PRE}. For the linear case, $S_1=b\rho$ and $\VEC S_2=b\VEC J$, and explicit expressions for the transport coefficients are given in the Supplemental Material. 

In the limit of short memory, the dynamics simplify considerably. When $\tau\to0$, $D_{12}$  and $\mu_{12}$ vanish, as shown in detail in Ref.~\cite{Mayo-Soto-PRE}.
This implies that $\rho$ decouples from $\rho_X$, and the KS equations are recovered for the density field, with $D=D_{11}$ and $\mu=\mu_{11}$.

\textit{Analysis.}
The simplest case to consider is the response to a static and stationary signal $\nabla l_0$. In this case, to dominant order $\rho_X$ is quadratic in $l_0$, with $\langle X \rangle\equiv \rho_X/\rho = -g_7 |\nabla l|^2$,  and the linear bacterial current is simply $\VEC J = \mu_{11} \nabla l_0$. Hence, $\mu_{11}$ is directly the stationary mobility. 

\begin{figure}[t!]
\begin{center}
\includegraphics[width=\columnwidth]{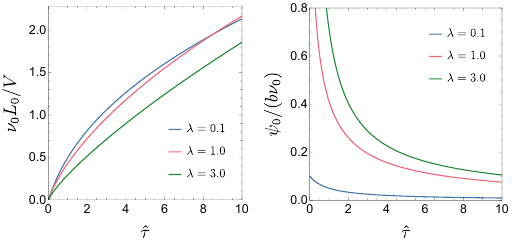}
\end{center}
\caption{Smoothing length $L_0$  (left) and  amplitude $\psi_0$ (right) of the static response function as a function of  $\hat\tau$ for the linear model, with $\alpha_1=0$.
Numerical solution for the coefficients truncating the expansion up to $n = 10$ terms. 
}
\label{fig.k0psi0}
\end{figure}

Increasing complexity, the next regime to consider is that of a stationary inhomogeneous chemotactic signal, $l(\VEC r)$. The linear KS expression for the flux \eqref{eq.ks} indicates that in equilibrium, when the fluxes vanish, the stationary concentration is  
$\rho(\VEC r) = \rho_0 e^{\mu l(\VEC r)/D}$, which is a local response. Upon linearization this gives $\rho(\VEC r) = \rho_0 + \rho_0\mu l(\VEC r)/D$. Linearizing Eqs.~\eqref{complet_rho_eq} and \eqref{complet_rhox_eq}, for a Fourier mode signal $l(\VEC{r},t)= l_0 + \eta l_1 e^{i\VEC{k}\cdot\VEC{r}}$ with $\eta\ll1$, the resulting densities are  $\rho =\rho_0 + \eta\psi_\rho l_1 e^{i\VEC{k}\cdot\VEC{r}}$ and $\rho_X =\eta \psi_{X} l_1e^{i\VEC{k}\cdot\VEC{r}}$, with the response functions
\begin{align} \label{eq.responselonretz}
\Psi_\rho(k,0)&=  \frac{\psi_0}{ \left[1+\left({k}/{k_0}\right)^2\right]}+\psi_1, \\
\Psi_X(k,0)&= -\frac{(D_{11}\psi_0/D_{12}) (k/k_0)^2}{ \left[1+\left({k}/{k_0}\right)^2\right]},
\end{align} 
with $\psi_0=\frac{D_{12}(D_{11}\mu_{21}-D_{12}\mu_{11})}{D_{11}(D_{11}D_{22}-D_{12}^2)}$, $\psi_1=\frac{D_{22}\mu_{11}-D_{12}\mu_{21}}{(D_{11}D_{22}-D_{12}^2)}$, and $k_0 =  \sqrt{\frac{D_{11}\gamma_1}{\tau \left(D_{11}D_{22}-D_{12}^2\right)}}$.
The density response function is Lorentzian in Fourier space, meaning that in real space is nonlocal, with a characteristic smoothing length $L_0=k_0^{-1}$. Figure \ref{fig.k0psi0} displays the  response amplitude $\psi_0$ and $L_0$, where it is evident that the latter grows with memory.

\begin{figure}[t!]
\includegraphics[width=.9\columnwidth]{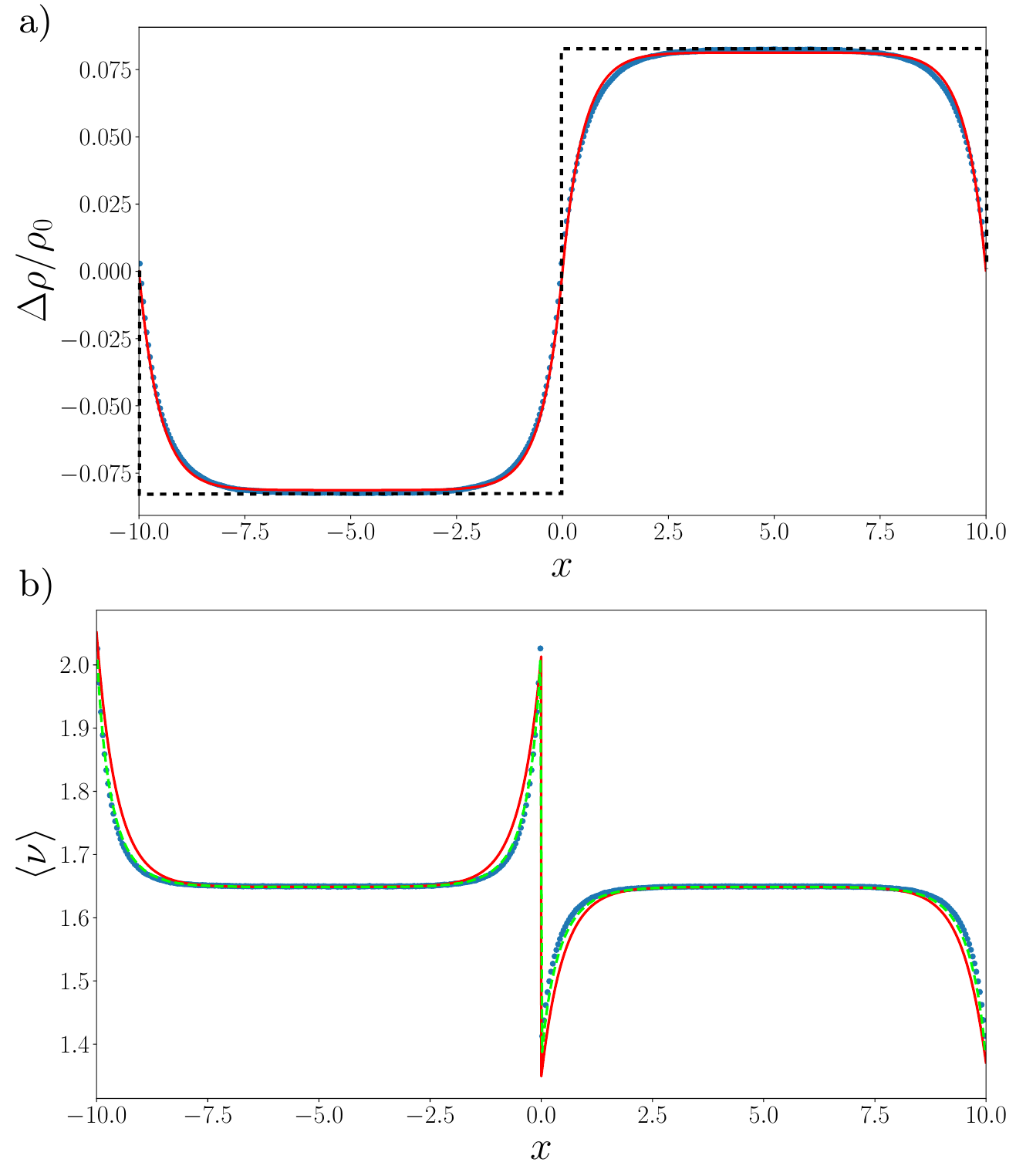}
\caption{Stationary normalized density (a) and average tumbling rate (b) profiles generated by a step function ligand signal in the linear model. The blue circles represent the results of the simulations of particles with $\hat\tau=1.0$, $\lambda=1.0$, and $\alpha_1=0$,  in a periodic square box of size $L = 20$, with a chemotactic signal of amplitude $l_1 = 0.4$. For thse parameters, $L_0=0.43$. In (a), the red solid line is the theoretical prediction and the solid dashed line the prediction of the KS model. 
In (b) the solid lines are the theoretical prediction for $\langle\nu\rangle$  by using $\langle X \rangle=\rho/\rho_X$ with the theoretical expression for $\rho(x)$ and $\rho_X(x)$ (red) or their measured values (green). 
Units are such that $V=\nu_0=b=1$.}
\label{rho_simulation}
\end{figure}

A possible experimental realization to measure this nonlocal effect consists on placing a chemotactic signal with a sharp step discontinuity, $l(x) = l_0 + l_1\operatorname{sgn}(x)$, with $\operatorname{sgn}(x) \equiv x/|x|$ the sign function. In experiments, $\rho_X$ is not directly accessible but the local tumbling rate is~\cite{khan2004fast,junot2022run}. In the linear model $X$ is approximately a Gaussian random variable of unit variance. Employing the cumulant expansion, one obtains
 $\langle\nu \rangle (\VEC{r}) =   \big( \nu_0 e^{\lambda^2/2} \big) e^{\lambda \langle X\rangle(\VEC{r})}$.
Computing the convolutions with \eqref{eq.responselonretz}, it results
$\rho(x)=\rho_0+\rho_0  l_1 \operatorname{sgn}(x)\left[\psi_1+\psi_0\left(1-e^{-k_0|x|}\right)\right]$ and 
$\rho_X(x)=-\rho_0 \frac{D_{11}}{D_{12}}  \psi_0  l_1 \operatorname{sgn}(x) e^{-k_0\left|  x\right|}$,
where it is evident the effect of the smoothing length. Figure \ref{rho_simulation} presents these profiles, which are compared with agent-based simulations of swimmers using the method described in Refs.~\cite{villa2020run,villa2023kinetic}

\begin{figure}[t!]
\begin{center}
\includegraphics[width=.9\columnwidth]{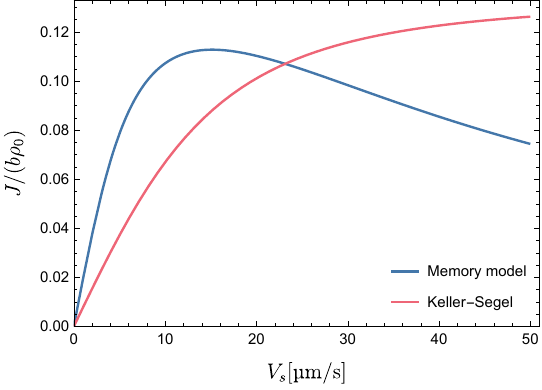}
\end{center}
\caption{Normalized bacterial current as a response to a traveling chemotactic wave of speed $V_s$, computed using the values of the linear model for \textit{E.\ coli}, for a periodic box of length $L=\SI{800}{\micro\meter}$, meaning that $\Psi_\rho$ is evaluated at $k=2\pi/L$. In blue the prediction of this letter and in red the prediction of the KS model (divided by 5 to help the comparison).}
\label{fig.current}
\end{figure}

Finally, we consider the case of a traveling chemotactic wave $l(x,t)=l_0 e^{i k(x-V_st)}$. Recent experiments performed with \textit{E.\ coli}, showed that, as a result of finite memory, the resulting chemotactic current is not monotonic with the wave speed $V_s$, presenting a maximum at $V_s\approx\SI{8}{\micro\meter/\second}$~\cite{li2017barrier}. Substituting in Eqs.~\eqref{complet_rho_eq} and \eqref{complet_rhox_eq}, gives for the linear response of the current that $J = V_s \Psi_\rho(k,V_sk) l_0$, where $\Psi_\rho(k,\omega)$ is the spatio-temporal density response function, with an  expression that  is direct to obtain, given explicitly in the Supplementary Material and Ref.~\cite{Mayo-Soto-PRE}. Figure~\ref{fig.current} shows the predicted current, which is compared with the equivalent prediction of the KS model, where the values of the linear model for \textit{E.\ coli} are used (see below). Although the experiment is performed in a strong non-linear response regime, it is remarkable that the present model predicts the existence of the maximum for a velocity in the same order of magnitude. 

\textit{Numerical values for E.\ coli.}
Using the parameters measures for the linear model given above, 
it is possible to give explicit expressions for the transport coefficients for  \textit{E.\ coli}. The only remaining parameter is $b$, which depends on the specific ligand to be considered. 
The diffusion coefficients are $D_{11}=\SI{1.3E3}{\micro\meter^2/\second}$,  $D_{12}=D_{21}=\SI{0.81E3}{\micro\meter^2/\second}$,  and  $D_{22}=\SI{0.99E3}{\micro\meter^2/\second}$. The chemotactic mobilities are $\mu_{11}/b = \SI{0.42E2}{\micro\meter^2/\second^2}$, $\mu_{12}/b = \SI{2.2E2}{\micro\meter^2/\second^2}$, $\mu_{21}/b = \SI{0.52E2}{\micro\meter^2/\second^2}$, and $\mu_{22}/b = \SI{2.6E2}{\micro\meter^2/\second^2}$.  Finally, the amplitude and smoothing length of the static response function are $\psi_0/b=\SI{0.032}{\second^{-1}}$ and $L_0=\SI{1.7E2}{\micro\meter}$, respectively. 

\textit{Conclusions.}
Bacterial chemotaxis is usually described by the macroscopic KS equations, which propose a local relation between the chemical gradient and the bacterial flux. However, this model does not consider the long memory present in the protein pathway that controls the tumbling rate and hence chemotaxis. This memory gives rise to intrinsic temporal and length scales, and consequently the KS equations are not valid when the chemical signal varies on these scales.

Using a kinetic approach, we derived the  equations for the relevant fields, which we show are the bacterial density and the internal protein concentration field. They take the form of reaction-diffusion equations with fluxes and source terms that depend on the spatio-temporal gradients of the chemoattractant field. Consistent with the inclusion of memory effects, the response is nonlocal, with a characteristic smoothing length. The associated transport coefficients, which replace the usual diffusion coefficient and chemotactic mobility, were obtained for a simple model of bacterial motion, but can be measured for specific systems or computed for other models~\cite{tu2008modeling,tu2013quantitative}. 

For a strain of \textit{E.\ coli}, it is possible to compute explicitly the transport coefficients and derive the associated temporal and length scales of the nonlocal response, which are $\SI{19}{\second}$ and $\SI{170}{\micro\meter}$, respectively. 
These values, which can be slightly smaller due to rotational diffusion, are of the order to those found in several experimental and natural conditions, showing the relevance of the derived equations. 
The proposed setup could allow to determine this length also for other bacteria.

The hydrodynamic equations~\eqref{complet_rho_eq} and \eqref{complet_rhox_eq}  need to be complemented with appropriate boundary conditions, where the natural choice is to use non-flux boundary conditions. However, as bacteria tend to accumulate at surfaces~\cite{berke2008hydrodynamic}, the problem is far from trivial and more research is needed to describe the interaction of bacteria with boundaries, considering memory effects.

\begin{acknowledgments}
This research is supported by Fondecyt Grants No.~1220536 (RS) and ANID Millennium Science Initiative Program NCN19$\_$170, Chile. MM acknowledges financial support by grant ProyExcel\_00505 funded by Junta de Andalucía and grant PID2021-126348NB-I00 funded by MCIN/AEI/10.13039/501100011033/ and ERDF ``A way of making Europe''.
\end{acknowledgments}


%

\clearpage
\appendix
\onecolumngrid

\section{SUPPLEMENTARY MATERIAL}

\subsection{Linear equations for the functions in $f^{(1)}$}
To first order in $\varepsilon$ the distribution function  $f^{(1)}$ satisfies the integro-differential equation
\begin{align}\label{3.29}
{ \left(   \frac{\gamma_1}{\tau}\rho_X \pdv{}{\rho_X} +   \mathcal{L}_0 \right)} f^{(1)} 
  =              
  \frac{1}{ \tau}  \left(\rho  E_2 +  \rho_X  E_4  \right)   \phi \pdv{l}{t} 
  + V \n \cdot \bigg[    \nabla   \rho  +
  u_1  \nabla \rho_{X}     
  +  \frac{1}{\tau}\left(\rho E_1+  \rho_X E_3 \right) \nabla l   \bigg] \phi,
\end{align} 
where the operator $\mathcal{L}_0$ is given by the right hand side of Eq.~(3) of the main text, with no ligand coupling ($B=0$), and 
\begin{subequations}\label{eqs.definicionesaux}
    \begin{align} 
E_1(X) &\equiv     A(X)B(X) -B^\prime(X), \\
E_2(X) &\equiv  E_1(X)- g_1 u_1(X)      , \\
E_3(X) &\equiv  u_1(X)E_1(X)  - u^\prime_1(X) B(X), \\
E_4(X)  &\equiv E_3(X)  - g_2 u_1(X)  .
 \label{2}
    \end{align}
\end{subequations}

A  series expansion  $M(X)=\sum_m M_m U_m(X)$, and similarly for $N,O,P,Q$, and $R$  is done for the functions that appear in the expression for $f^{(1)}$ [Eq.~(7) of the main text]. Substituting this expansion in Eq.~\eqref{3.29}, the coefficients can be obtained thanks to the the orthogonality of $U_m$ (see Ref.~\cite{Mayo-Soto-PRE} for details). 
It results that
\begin{subequations}\label{eqsalgM-Q}
    \begin{align} 
  &  M_n =  -\frac{\nu_0  \tau \Omega_d}{\gamma_n} \int d X  u_m E_2 \phi , \label{4.33d} \\
  & N_n =  -\frac{\nu_0  \tau \Omega_d}{(\gamma_n-\gamma_1)} \int d X  u_n E_4 \phi, \label{4.33p} \\       
 & \sum_m c_{mn1}O_m  =  - \nu_0 \tau  \delta_{n0},  \label{4.31}\\
 &  \sum_m c_{mn1}P_m  =  - \nu_0 \tau  \delta_{n1}, \label{4.33a} \\
 &    \sum_m c_{mn1} Q_m = -  \nu_0  \tau  \Omega_d \int d X  u_n  E_1  \phi, \label{4.33b} \\
 &  \sum_m c_{mn1}  R_m - \gamma_1 R_n=  -\nu_0  \tau  \Omega_d \int d X  u_n  E_3 \phi. 
    \end{align}
\end{subequations}

Here, 
\begin{align}
c_{mn1}=\nu_0 \tau b_{mn}(1-\alpha_1)+ \gamma_m\delta_{mn}, \label{eq.defc}
\end{align}
with 
\begin{align}
\label{matrixbmm}
b_{mm'} & = \Omega_d\int_{-\infty}^{\infty}  d X \phi^{-1} C(X) U_m(X) U_{m^\prime}(X).
\end{align}

\subsection{General expressions for the transport coefficients}

\begin{subequations}
\begin{align}
D_{11} &=  -\frac{V^2 O_0 }{d \nu_0}  ,\label{eq9.1A}\\
D_{12} &=\frac{V^2 P_0}{d\nu_0},\label{eq9.2A}\\
 \mu_{11} &=\frac{V^2 Q_0}{d\hat\tau} ,\label{eq9.3A}\\
 \mu_{12} &=-\frac{V^2 R_0}{d\hat\tau }.\label{eq9.4A}\\
 D_{21} &=  \frac{V^2 O_1 }{d \nu_0}  ,\label{eq9.1B}\\
D_{22} &=-\frac{V^2 P_1 }{d\nu_0},\label{eq9.2B}\\
 \mu_{21} &=-\frac{V^2 Q_1}{d\hat\tau} ,\label{eq9.3B}\\
 \mu_{22} &=\frac{V^2 R_1}{d\hat\tau }.\label{eq9.4B}
\end{align} 
\end{subequations}

\begin{subequations} \label{eq.gi}
\begin{align}
g_1 &=\Omega_d \int dX u_1'  B\phi ,\\ 
g_2 &=\Omega_d \int dX u_1'  u_1 B\phi,  \\
g_3&=  \frac{1}{ \nu_0\tau} \int  dX  u_1^\prime BM, \\
g_4&=  \frac{1}{ \nu_0\tau} \int  dX u_1^\prime BN, \\ 
g_5 &=  \frac{V^2}{ d\nu_0} \int  dX u_1^\prime BO, \\ 
g_6 &=  \frac{V^2}{ d \nu_0} \int  dX  u_1^\prime BP, \\
g_7 &=  \frac{V^2}{ d \nu_0\tau} \int  dX u_1^\prime BQ, \\
g_8 &=  \frac{V^2}{ d \nu_0\tau} \int  dX u_1^\prime BR. 
\end{align}
\end{subequations}

\subsection{Intermediate results for the linear model ($A(X)=X$, $B(X)=b$, $C(X)=e^{\lambda X}$)}

In the linear model, the stationary solution and eigenfunctions are:
$\phi  = \frac{e^{-X^2/2}}{\Omega_d \sqrt{2\pi}}$, $u_n(X)=H_n(X/\sqrt{2})/\sqrt{ 2^n n!}$, with $H_n$  the Hermite polynomial of order $n$ [such that $H_0(x) = 1$, $H_1(x) = 2x$,  . . . ]\cite{arfken2011mathematical}, and $\gamma_n=n$~\cite{villa2023kinetic}.

Noting that in this case $u_1'=H_0(X/\sqrt{2})=1$, Eqs.~\eqref{eq.gi}a-b imply that $g_1=b$ and $g_2=0$. Hence, by Eqs.~\eqref{eqs.definicionesaux}, $E_1(X)=bX$, $E_2(X)=0$ and $E_3(X)=E_4(X)= b(X^2-1)$. Finally, the orthogonality of the Hermite polynomials imply that
$g_3=g_4=0$, $g_5=bV^2 O_0/(d\nu_0)=-bD_{11}$, $g_6=bV^2 P_0/(d\nu_0)=bD_{12}$, $g_7=bV^2 Q_0/(d\hat\tau)=b\mu_{11}$, and $g_8=bV^2 R_0/(d\hat\tau)=-b\mu_{12}$

In the linear model, the matrix elements \eqref{matrixbmm} are
\begin{align}
\label{matrixbmm_linear}
b_{mm'}(\lambda)  =&\frac{1}{\sqrt{2\pi 2^{m+m'} m! m'!}}  \int_{-\infty}^{\infty} e^{\lambda X-X^2 / 2} H_{m}(X / \sqrt{2}) H_{m'}(X / \sqrt{2}) d X, \nonumber \\
 =&e^{\lambda^2 / 2} \begin{pmatrix}
1 &  \lambda & \cdots \\
 \lambda & \left(1+\lambda^2\right) & \cdots \\
\vdots & \vdots & \ddots
\end{pmatrix}.
\end{align}

\subsection{Analytical solution of the linear equations for the linear model}

Keeping up to $n=1$, that is, considering two polynomials, the solution of Eqs.~\eqref{eqsalgM-Q} for the linear model is
\begin{subequations}
\begin{align} \label{3.46}
O_0  =& -\frac{1}{e^{\frac{\lambda^2}{2}}}\left[\frac{1}{1-\alpha_1}+\frac{\lambda^2 \hat\tau e^{\frac{\lambda^2}{2}}}{1+(1-\alpha_1)e^{\frac{\lambda^2}{2}} \hat\tau}\right],\\
O_1  =& \frac{\lambda \hat\tau}{1+(1-\alpha_1) e^{\frac{\lambda^2}{2}} \hat\tau},\\
P_0 =& Q_0/b  = \frac{\lambda \hat\tau}{1+(1-\alpha_1) e^{\frac{\lambda^2}{2}} \hat\tau},\\
P_1  =& Q_1/b  = -\frac{\hat\tau}{ 1+(1-\alpha_1) e^{\frac{\lambda^2}{2}} \hat\tau},\\
R_0 =& R_1 =0.
\end{align}
\end{subequations}
where the expression \eqref{matrixbmm_linear} for the $b_{mm'}$ matrix was used.

If we truncate up to $n=2$,  that is, considering three polynomials, the results are
\begin{subequations}
\label{eqM.orden2}
\begin{align} \label{3.47}
                  O_0  =&-\frac{e^{-\frac{\lambda^2}{2}} \left[(1-\alpha_1)^2 e^{\lambda^2} \left(\lambda^4+2 \lambda^2+2\right) \hat\tau^2+(1-\alpha_1) e^{\frac{\lambda^2}{2}} \left(\lambda^4+8 \lambda^2+6\right) \hat\tau+4\right]}{2(1-\alpha_1) \left[(1-\alpha_1)^2 e^{\lambda^2} \hat\tau^2+(1-\alpha_1) e^{\frac{\lambda^2}{2}} \left(2 \lambda^2+3\right) \hat\tau+2\right]}, & \\ 
        O_1  = &{\frac{\lambda \hat\tau \left[ (1-\alpha_1)e^{\frac{\lambda^2}{2}} \left(\lambda^2+1\right) \hat\tau+2\right]}{\left[(1-\alpha_1)^2e^{\lambda^2} \hat\tau^2+(1-\alpha_1)e^{\frac{\lambda^2}{2}} \left(2 \lambda^2+3\right) \hat\tau+2\right]}},\\
        O_2  =&-{\frac{\lambda^2 \hat\tau \left[(1-\alpha_1)e^{\frac{\lambda^2}{2}} \hat\tau-1\right]}{\sqrt{2}\left[(1-\alpha_1)^2e^{\lambda^2} \hat\tau^2+(1-\alpha_1)e^{\frac{\lambda^2}{2}} \left(2 \lambda^2+3\right) \hat\tau+2\right]}},
\end{align}
\end{subequations}
\begin{subequations}
\label{eqN.orden2}
\begin{align}
                  P_0  =Q_0/b = &{\frac{\lambda \hat\tau \left[(1-\alpha_1)e^{\frac{\lambda^2}{2}} \left(\lambda^2+1\right) \hat\tau+2\right]}{(1-\alpha_1)^2e^{\lambda^2} \hat\tau^2+(1-\alpha_1)e^{\frac{\lambda^2}{2}} \left(2 \lambda^2+3\right) \hat\tau+2}}, \\ 
        P_1  =Q_1/b = &-{\frac{\hat\tau \left[(1-\alpha_1)e^{\frac{\lambda^2}{2}} \left(2 \lambda^2+1\right) \hat\tau+2\right]}{\left[(1-\alpha_1)^2e^{\lambda^2} \hat\tau^2+(1-\alpha_1)e^{\frac{\lambda^2}{2}} \left(2 \lambda^2+3\right) \hat\tau+2\right]} },\\
        P_2  =Q_2/b =& {\frac{\sqrt{2}(1-\alpha_1)e^{\frac{\lambda^2}{2}} \lambda \hat\tau^2}{(1-\alpha_1)^2 2 e^{\lambda^2} \hat\tau^2+(1-\alpha_1)2 e^{\frac{\lambda^2}{2}} \left(2 \lambda^2+3\right) \hat\tau+4}},
    \end{align}
\end{subequations}

\begin{subequations}
\label{eqP.orden2}
\begin{align}                  
{R_0/b  =} &\frac{2 (1-\alpha_1 ) \lambda ^2 e^{\frac{\lambda ^2}{2}} \hat\tau ^2}{2 \left(\lambda ^2+1\right)+(1-\alpha_1 ) e^{\frac{\lambda ^2}{2}} \left(\lambda ^2+2\right) \lambda ^2 \hat\tau -2 (1-\alpha_1 )^2 e^{\lambda ^2} \hat\tau ^2}  , \\ 
       { R_1/b  =} &-\frac{ 2\lambda  \hat\tau  \left[\lambda ^2+2-2 (1-\alpha_1) e^{\frac{\lambda ^2}{2}} \hat\tau \right]}{2 (1-\alpha_1)^2 e^{\lambda ^2} \hat\tau ^2-(1-\alpha_1 ) e^{\frac{\lambda ^2}{2}} \left(\lambda ^2+2\right) \lambda ^2 \hat\tau -2 \left(\lambda ^2+1\right)},\\
       { R_2/b  =}& \frac{2\sqrt{2}\hat\tau  \left[\lambda ^2+1-(1-\alpha_1 ) e^{\frac{\lambda ^2}{2}} \hat\tau \right]}{2 (1-\alpha_1 )^2 e^{\lambda ^2} \hat\tau ^2-(1-\alpha_1 ) e^{\frac{\lambda ^2}{2}} \left(\lambda ^2+2\right) \lambda ^2 \hat\tau -2 \left(\lambda ^2+1\right)} .
    \end{align}
    \end{subequations}

\subsection{Linear response function}\label{expressions.gral}

The response functions are obtained by linearizing the hydrodynamic equations and solving them in Fourier space. The result is
\begin{align}
\Psi_\rho(k,\omega) &= \frac{k^2 [\gamma_1 \mu_{11} + k^2 (D_{22} \mu_{11} - D_{12} \mu_{21}) \tau - 
   i (D_{12} g_1 + \mu_{11}) \tau \omega]}{
D_{11} \gamma_1 k^2 + (-D_{12}^2 + D_{11} D_{22}) k^4 \tau - 
 i [\gamma_1 + (D_{11} + D_{22}) k^2 \tau] \omega - \tau \omega^2}.
\end{align}

\end{document}